\documentclass[aps, prb, a4paper, reprint, 10pt, showpacs, twocolumn, groupedaddress, longbibliography]{revtex4-2}
\usepackage{graphicx}
\usepackage{fullpage}
\usepackage{amsmath}
\usepackage{amssymb}
\usepackage{bm}
\usepackage{hyperref}
\usepackage{natbib}
\usepackage{inputenc}

\let\cref\ref

\begin{document}
  
\title{Synchronization of Bloch oscillations by gate voltage modulation }
\author{Janis Erdmanis} \author{Yuli Nazarov}
\affiliation{Kavli Institute of Nanoscience, Delft University of Technology, 2628 CJ Delft, The Netherlands} 

\begin{abstract}
We propose to synchronize Bloch oscillations in a double phase-slip junction by modulating the gate voltage rather than the bias voltage. We show this is advantageous and the relatively small a.c. modulation of the gate voltage gives rise to the pronounced  plateaux of quantized current of the width of the order of Coulomb blockade threshold. 

We theoretically investigate the setup distinguishing three regimes of the strong, weak, and intermediate coupling defined by the  ratio of the gate capacitance $C$ and the effective capacitance of the phase-slip junctions. An important feature of the intermediate coupling regime is the occurrence of the fractional plateaux of the quantized current. We investigate the finite temperature effects finding an empirical scaling for the smoothing of integer plateaux.

\end{abstract}

\maketitle
\thispagestyle{empty}

\section{Introduction}

An elementary process giving rise to resistance in a quasi-one-dimensional superconducting wire is a phase-slip: an event where the magnitude of the superconducting order parameter locally and momentarily reaches  to zero, allowing the phase difference of the superconducting order parameters on the left and on the right to slip by  $2 \pi$. \cite{McCumber1970,Langer1967} see \cite{Arutyunov2008} for review. This corresponds to a magnetic flux quantum $\Phi_0$ moving across the wire, or, equivalently, a voltage pulse with $\int dt V(t) = \Phi_0$. The phase slips due to thermal and quantum fluctuations have been observed and identified in course of thorough and difficult experiments \cite{Lau2001,Aref2012,Voss2021}.

The quantum coherence between individual phase slip events results in modification of the ground state of superconducting systems that become a superposition of the states \cite{Mooij2005} that differ in flux quanta. Quantum coherent phase-slip states have been observed in superconducting nanowires\cite{Astafiev2012,Peltonen2013} and the chains of Josephson junctions that in many respects are similar to the superconducting nanowires \cite{Pop2010}. There is a remarkable duality between the coherent tunnelling of flux quanta due to the phase slips and  the coherent tunnelling of Cooper pairs transferring charge $2e$ that is the base of Josephson effect \cite{Mooij2006}, so that each Josephson electronic circuit has a dual counterpart made from phase-slip junctions, and vice versa. This gave rise to many theoretical suggestions and experimental work. \cite{Arutyunov2008, Vanevi2012, Hriscu2011, Hriscu2011a, Hriscu2013,DiMarco2015,Rastelli2013,Corlevi2006,Manucharyan2012, Houzet2020,Kuzmin2021}  

A phase-slip junction in a high resistive environment gives rise to a Coulomb blockade of the current up to a voltage threshold, that is dual to a zero-voltage state up to a current threshold in a Josephson junction. 

It has been predicted \cite{Averin1985, Schoen1990,Geigenmueller1988} that this should give rise to an effect that is potentially indispensable for applications: synchronization of Bloch oscillations. Biasing a phase-slip junction with combination of d.c. and a.c. bias should give rise to current plateaux with the value corresponding to the a.c. frequency $\omega$: $I = (e/\pi)\omega$. This may enable a high-precision current standard that is dual to the Josephson voltage standard \cite{Mooij2006}.    
However, it is much more difficult to realize a high-impedance environment than the low-impedance one, and prevent overheating of such environment by power dissipation. Although the phase-slip Coulomb blockade feature has been reliably observed,
(e.g. \cite{Constantino2018,Fenton2016}), the attempts to achieve the synchronization of Bloch oscillations \cite{Kuzmin1991, Corlevi2006, Lehtinen2012, Webster2013} have not yet demonstrated a precision even remotely comparable to  a Josephson voltage standard \cite{Wang2019,Pekola2013}. 

Recently, much experimental and theoretical interest was received by a double phase-slip junction  \cite{Hriscu2011a, Hongisto2012, Tobias2019}. The total phase-slip amplitude there is a result of interference of the phase-slips in constituent junctions, this interference being affected by a gate voltage supplied via a capacitive coupling. The gate charge dependence has been successfully demonstrated in spectroscopy measurements of the phase-slip qubit level positions  \cite{Graaf2018}and in the measurements of the Coulomb blockade threshold \cite{Hongisto2012,Arutyunov2017,Graaf2018}.

In this Article, we propose synchronization of Bloch oscillations in a double phase-slip junction by the a.c. gate voltage and theoretically investigate this phenomenon in a variety of regimes. 

Let us explain here why, in our opinion, the gate voltage synchronization is advantageous in comparison with the standard bias synchronization. The reason is general although rather technical, at least from the theoretical point of view. A phase-slip junction should be embedded to a high-impedance environment with substantial capacitance. While this capacitance is irrelevant for d.c. bias, there is an overwhelming $RC$ filtering of the a.c. signal. To get a substantial a.c. signal at the junction, one needs to increase a.c. bias by orders of magnitude to compensate for the filtering. This a.c. bias leads to substantial dissipation in the environment, its overheating and destruction of the synchronization by the thermal noise generated by this overheating.
This is likely explanation of the fact that a prominent Coulomb blockade feature at d.c. bias does not give rise to high-quality synchronization if a.c. bias is applied. In contrast to this, the a.c. gate voltage signal propagates in low-resistance environment and should provide much less dissipation.

In this Article, we systematically analyse the phenomenon in the quasiclassical limit corresponding to the limit of high impedance mostly concentrating on the peculiarities of current plateaux.
 We distinguish three regimes. (i) A strong coupling regime corresponds to the limit of small gate capacitance $C$ as compared to effective phase-slip junction capacitances. Most experimental setups are in this regime \cite{Hongisto2012,Arutyunov2017,Graaf2018}. The harmonic gate voltage modulation gives rise to multiple {\it integer} plateaux. We show that in this regime the width of the plateau can be made comparable with the Coulomb blockade threshold voltage even for very asymmetric junctions at sufficiently large modulation amplitudes. (ii) An intermediate coupling regime where the effects of finite gate capacitance are essential. We demonstrate the appearance of {\it fractional} plateaux with the width $\propto C$. This is a dual of half-integer Shapiro steps observed in Josephson SQUIDS \cite{Romeo2004}.(ii) A weak coupling regime of big $C_g$ where the gate capacitor plays the role of an effective d.c. voltage source and efficiently decouples. 

 For all three regimes, we analyze the effect of the thermal noise on the plateaux. We demonstrate that that integer plateaux vanish at the  temperature $k_B T \approx 0.06 e \Delta V$, $\Delta V$ being the plateau width in the absence of noise.

The paper is organized as follows. In the Section \cref{sec:description-devices--1} we give the description of the setup and the system of equations governing its dynamics. We address the stationary regime in Section \ref{sec:sign-modul-phase}. We discuss the emergence of current plateaux in strong (Section \ref{sec:strong-coupl-regime}), intermediate (Section \ref{sec:fractional}), and weak (Section \ref{sec:weak-coupling-regime})coupling regimes. 
 In Section \ref{sec:finite-temp-effects} we address the finite temperature effects in all three regimes. We conclude in Section \cref{sec:conclusion}.

\section{The setup}\label{sec:description-devices--1}

\begin{figure}[ht]
  \centering
  \includegraphics[width=0.8\columnwidth]{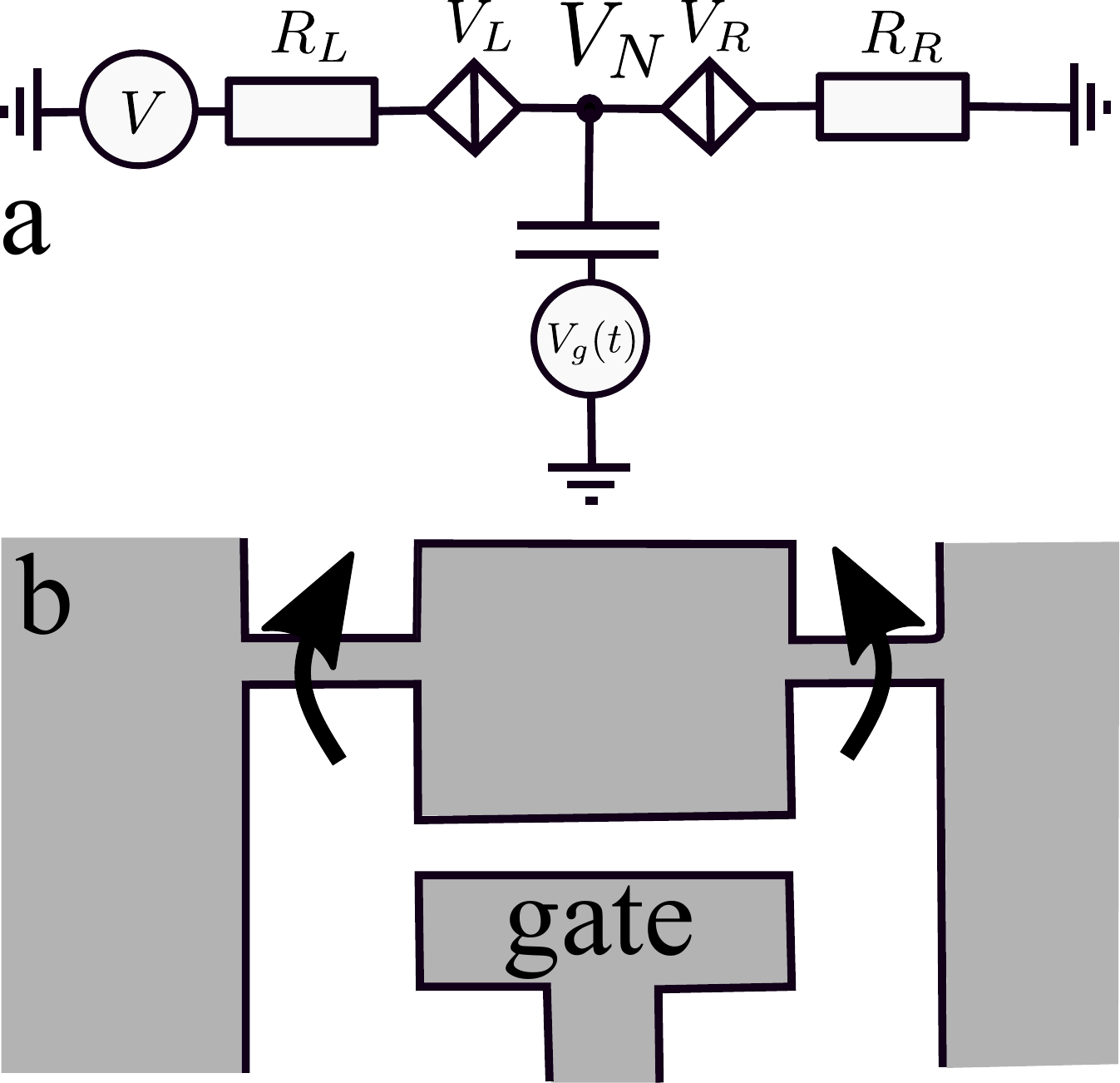}

  \caption{
Two phase slip junctions in series in resistive environment coupled to a capacitor in between where the gate voltage is applied.}
\label{fig:setup}
\end{figure}

In this Article, we address a double phase-slip junction setup (Fig. \ref{fig:setup}). We require two phase slip junctions in series: those can be realized lithographically as short nanowire-type constrictions in a superconducting film of a high normal-state resistance\cite{Astafiev2012} (Fig.\ref{fig:setup}b). We do not have to impose any stringent conditions on homogeneity and regularity of the materials in use: we only need the phase slip tunnelling amplitudes via the narrow parts of the setup. The tunnelling amplitudes may be due to phase slips arising uniformly in the wires, or they can be dominated by the slips in the most narrow place of the wire, or even occur in a tunnel-type Josephson junction that is formed in the wire: this is not important, as explained in \cite{Vanevi2012} in more detail. Eventually, the nanowire could be a Josephson junction chain with the phase slip occurring at the weakest junction. Experimentally, the amplitude in a single junction, either left or right one, is determined from the observation of the Coulomb blockade threshold voltage, $V_{L,R}$ being the threshold voltages for the junctions of the setup. The junctions are embedded in the high-impedance environment. We took the simplest and most frequently used model of such environment: the frequency-independent resistors $R_{L,R}$ on both sides of the double junction. The important part of the setup is the 
gate electrode coupled via the capacitance $C$ to the node between the junctions.

We consider here the quasi-classical limit that is justified when the resistance exceeds by far the quantum unit, $R_{L,R} G_Q \gg 1$, $G_Q \equiv e^{2}/(\pi \hbar)$.  In this case, the charge passed though a junction is a good classical variable subject to little quantum fluctuations. We will use a dimensionless charge $q = Q\pi/e$. In these notations, each junction can be regarded as a non-linear capacitor with the voltage drop depending on $q$,  $V(q) = V_{L,R} \sin q_{L,R}$.

The voltage drops at the resistors are given by corresponding currents, $\frac{e}{\pi} R_{L,R} \dot q_{L,R}$. We assume the system is biased from the left with the voltage $V$. Introducing the voltage $V_N(t)$ at the node between the junctions, we equate the voltage drops as follows:
\begin{align}
\label{eq:junctions}
  V_N(t) &= V_R \sin q_R + \frac{e}{\pi} R_R \dot q_R \\
  V - V_N(t) &= V_L \sin q_L + \frac{e}{\pi} R_L \dot q_L
\end{align}

Each of these two equations is equivalent to a standard voltage-biased single-junction equation. The coupling between the junctions comes about the fact that the voltage at the node $V_N(t)$ also depends on the  charge accumulated at the capacitor $C$ . It is also contributed by the gate voltage, 
\begin{align}
\label{eq:gate}
  V_N(t) = \frac{e}{\pi} \frac{q_L - q_R}{C} + V_g(t)
\end{align}

In the weak coupling limit $C \to \infty$ the capacitor efficiently uncouples the phase-slip junctions \cref{sec:weak-coupling-regime}. In the opposite limit of strong coupling (\cref{sec:strong-coupl-regime}), the two phase-slip junctions effectively work as one with an amplitude that is the some of the two. The interference of two amplitudes  can be tuned by gate voltage.

Now we can account for the voltage noise coming from the thermal fluctuations in the large resistors. The noise gives a stochastic addition  $\xi(t)$ to the voltage drop at a resistor satisfying $\langle \xi(t) \xi(t') \rangle = 2 k_B T R \delta(t - t')$. The resulting system of equations which we analyze in Article thus reads:
\begin{equation}
  \begin{cases}
    \frac{e}{\pi} R_L \dot q_L =  V - V_L \sin q_L  - V_N(t) + \xi_L(t)\\
    \frac{e}{\pi} R_R \dot q_R = V_N(t) - V_R \sin q_R  + \xi_R(t)\\
    V_N(t) = \frac{e}{\pi}\frac{q_L - q_R}{C} + V_g(t) \label{eq:1}
  \end{cases}
\end{equation}
These are the stochastic evolution equations for two variables $q_R,q_L$.

Owing to the duality mentioned, a similar set of equations describes a Josephson junction system in a low-impedance environment. This system is a d.c. current-biased two-junction SQUID with an extra inductance in the loop  subject to a time-dependent flux penetrating the loop (a dual of the time-dependent gate voltage). Such Josephson circuits have been studied in \cite{Vanneste1988,Romeo2004} but, to our knowledge, have not been put into practice.

\section{Stationary regime} \label{sec:sign-modul-phase}

In this Section, we shortly describe and illustrate the stationary regime where both bias and gate voltage do not depend on time. Let us start with Colomb blockade when (in the absence of noise) no current is flowing till the bias voltage $V$ reaches a certain threshold value $V_{{\rm th}}$. 
In Coulomb blockade regime, $V = V_L \sin q_L +V_R \sin q_R$, and the charges $q_{L,R}$ are related by 
\begin{equation}
q_L - q_R = - q_g + \frac{\pi C V_R}{e} \sin q_R
\end{equation}
where we have introduced the gate-induced charge $q_g \equiv \pi C V_g/e $. The periodicity of $\sin q_{L,R}$ implies the periodicity of the results for $V_{{\rm th}}$ in $q_g$, that is, in gate voltage.

These results also depend on the ratio between the gate capacitance and effective capacitance of the phase-slip junctions.  We will use the dimensionless parameter 
\begin{align}
  \tilde C = \frac{\pi C (V_L + V_R)}{2e}
\end{align}
to characterize the ratio and distinguish the regimes. 

The strong coupling regime corresponds to small $\tilde{C}$. In this case, two phase-slip junctions are equivalent to a single junction. The difference $q_R-q_L$ is set to $q_g$, and the overall phase-slip amplitude is a sum of two amplitudes corresponding to tunneling in the junctions,
\begin{equation}
Am = V_L e^{i q_L} + V_{R} e^{i q_L + i q_g}
\end{equation}
The gate voltage controls the relative phase shift and thus the interference of two amplitudes. 
The threshold voltage is determined by the maximum modulus of this amplitude,
\begin{equation}
V_{{\rm th}} = \sqrt{V_R^2 +V_L^2 + 2V_R V_L \cos q_g} 
\end{equation}
The threshold vanishes at $V_L=V_R$, $q_g {\rm mod} 2\pi = \pi$ as the result of destructive interference. 

The large capacitance effectively decouples the phase slips in the junctions, so in the opposite regime of weak coupling $\tilde{C} \gg 1$ the threshold voltage is thus a sum of two thresholds plus a a small periodic correction:
\begin{equation}
\frac{V_{{\rm th}}}{V_L+V_R} = 1 - \frac{\bar{q}_g}{ 2 \tilde{C}},
\end{equation}
$\bar{q}_g \equiv (\pi C V_R/e - q_g) {\rm mod} 2\pi$.

\begin{figure*}[ht] 
  \centering
  \begin{center}
    \includegraphics[width=\columnwidth]{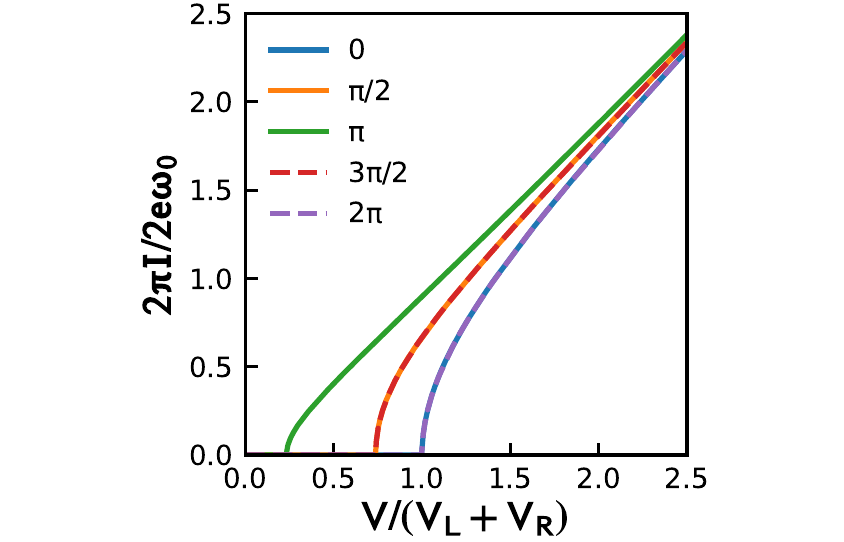}
    \includegraphics[width=\columnwidth]{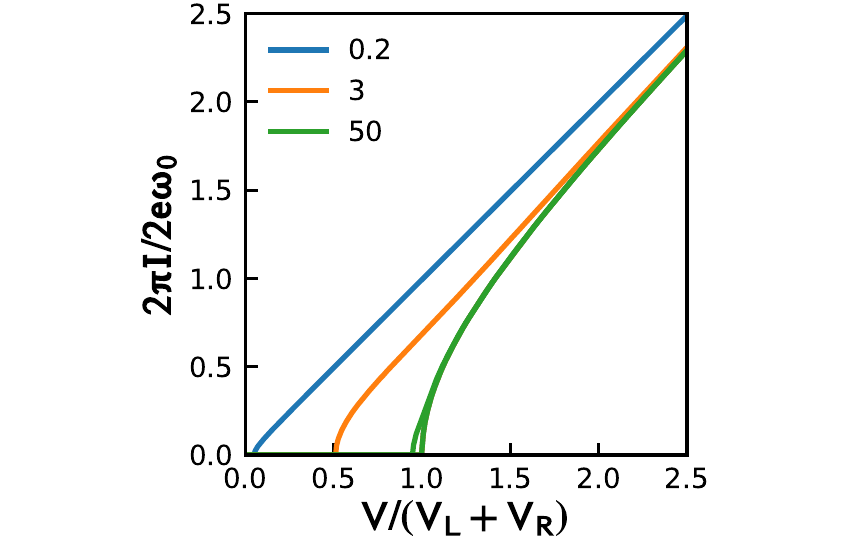}
  \end{center}

\caption{I-V characteristics in stationary regime. Symmetric setup, $R_L=R_R$, $V_L=V_R$. Left: Intermediate coupling, $\tilde C = 1$. The labels give the gate-induced charge $q_0$. The Coulomb blockade threshold is modulated by $q_0$, but does not vanish at $q_0 = \pi$ owing to finite $\tilde{C}$. Right:  I-V characteristics for $q_0=0,\pi$ and three different $\tilde{C}$ shown in the labels. The curves collapse at $q_0$, which is the peculiarity of the symmetric setup. The minimum threshold approaches the maximum one upon increasing $\tilde C$ manifesting the weak coupling regime.}\label{fig:blockade}
\end{figure*}

If the bias voltage exceeds the threshold, a d.c. current $I$ flows in the circuit. It is accompanied by a.c. (Bloch) oscillations with the frequency $\omega_B = \pi I/e$ corresponding to the d.c. current.  We obtain the I-V characteristics by solving the system \cref{eq:1} at given $V,V_g$ at a long time interval $(0,t)$ and calculating the time-averaged current from $I = (q_{L,R}(t) - q_{L,R}(0))/t$. The characteristics evaluated  are shown in the Figure \ref{fig:blockade}.

We see that I-V curves qualitatively follow the same shape typical for a single junction: a sharp square-root rise immediately after threshold and Ohmic behaviour $I = V/(R_R+R_L)$ at $V \gg V_{L,R}$. The threshold is however modulated by the gate voltage. We measure the current in units of $2e\omega_0$, $\omega_0$ being the frequency scale 
determined by the phase slips,
\begin{align}
  \omega_0 = \frac{\pi}{e} \frac{V_L + V_R}{R_L + R_R}
\end{align}
and change the definition of $q_g$ to compensate for the average $V_N$ in Ohmic regime, 
\begin{align}
   q_0 &= \frac{\pi C}{e} \left(
  \frac{R_R}{R_L + R_R} V - V_g 
   \right)
\end{align}
In the left pane of the Figure \ref{fig:blockade}we plot the $I-V$ characteristics for the intermediate coupling $\tilde{C}$ at several $q_0$. The threshold is modulated by the gate voltage but does not vanish even for the symmetric case considered since the finite capacitance effects suppress the destructive interference, its minimum value being $\approx 0.2 (V_L+V_R)$. In the right pane of the Figure \ref{fig:blockade} we plot the $I-V$ curves at various $\tilde{C}$ at $q_0= 0, \pi$.
At $q_0=0$ maximizing the threshold all characteristics are the same: this is a peculiarity of the symmetric case. At $q_0=\pi$ the threshold changes from almost zero to almost maximum value upon decreasing the capacitance and thus decoupling the junctions.

\section{Plateaux in strong coupling regime}\label{sec:strong-coupl-regime}

The most important application of the circuit is the synchronization of Bloch oscillations with an extra external a.c. signal of frequency $\omega$.
Without such signal, the solutions of the dynamical equations are degenerate with respect to the phase of the oscillations owing to time translation symmetry. The a.c. signal breaks the symmetry and the phase locks with that of the signal. In the d.c. measurement this is manifested as a current plateau: the d.c. current does not depend on the bias voltage in a certain interval of the voltages of the width $\Delta V$, and the value of the current is determined by the frequency only. In this Section and two following ones we explore the current plateaux in strong, intermediate and weak coupling regimes respectively. We consider the periodic modulation of the gate voltage, concentrating on harmomic one: $q_g(t) = q_0 + A \sin (\omega t)$.

\begin{figure}[ht]
  \centering
  \includegraphics[width=\columnwidth]{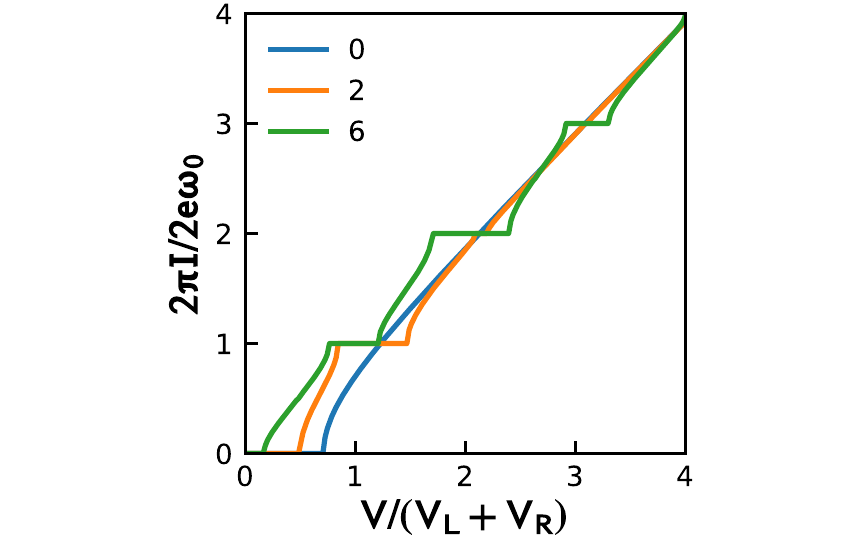}  
  \caption{I-V characteristics in the strong coupling regime, $\tilde{C} = 0.2$, for the different gate voltage amplitudes $A$ shown in the labels.  The setup is symmetric, $q_0=\pi/2$, and $\omega = \omega_0$.  Upon increasing the amplitude, the current plateaux develop at the multiples of the modulation frequency  $I_k = k e\omega/\pi$.  }\label{fig:strong}
\end{figure}

Some examples of $I-V$ characteristics are shown in Fig. \ref{fig:strong}. We observe the current plateaux at multiples of the modulation frequency  $I_k = k e\omega/\pi$. The width of the plateau $\simeq (V_L+V_R)$ at $A \simeq 1$. At small $A$, the width of the k-th plateau scales as $A^k$. The width is a non-monotonous function of $A$.

\begin{figure*}[ht]
  \centering
  \includegraphics[width=0.24\textwidth]{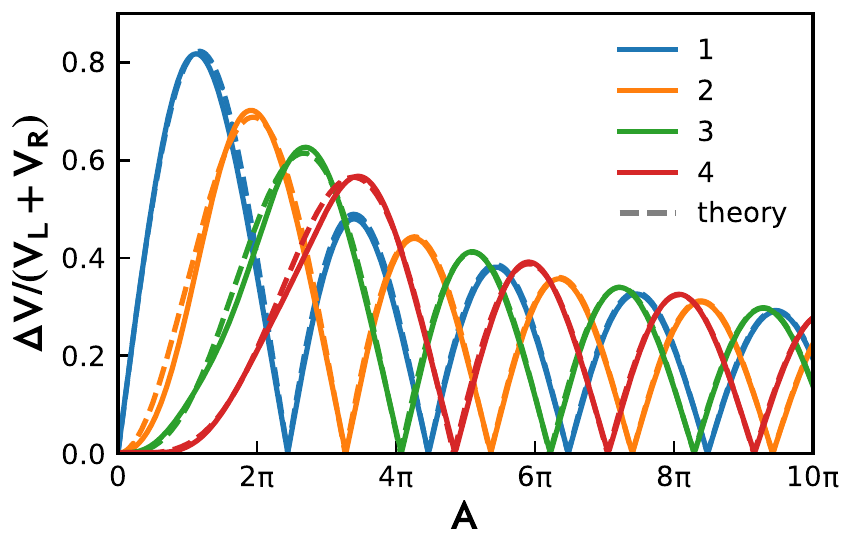}
  \includegraphics[width=0.24\textwidth]{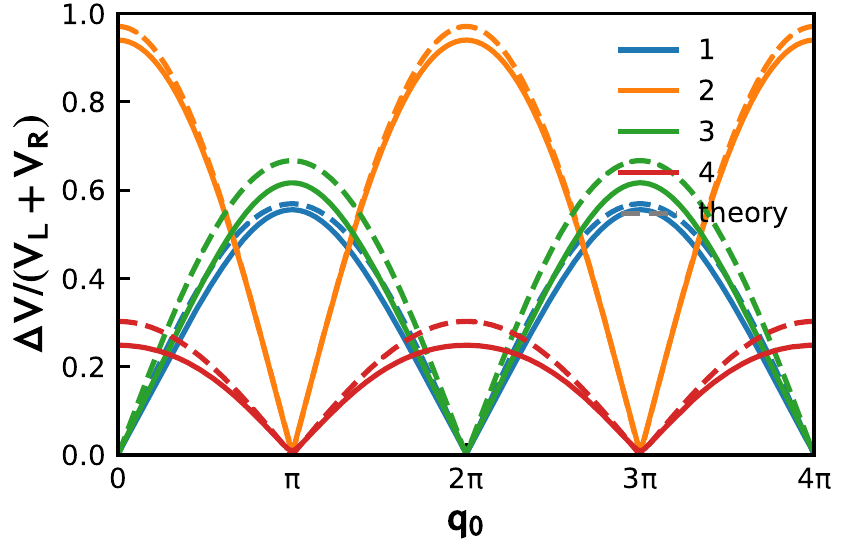}
  \includegraphics[width=0.24\textwidth]{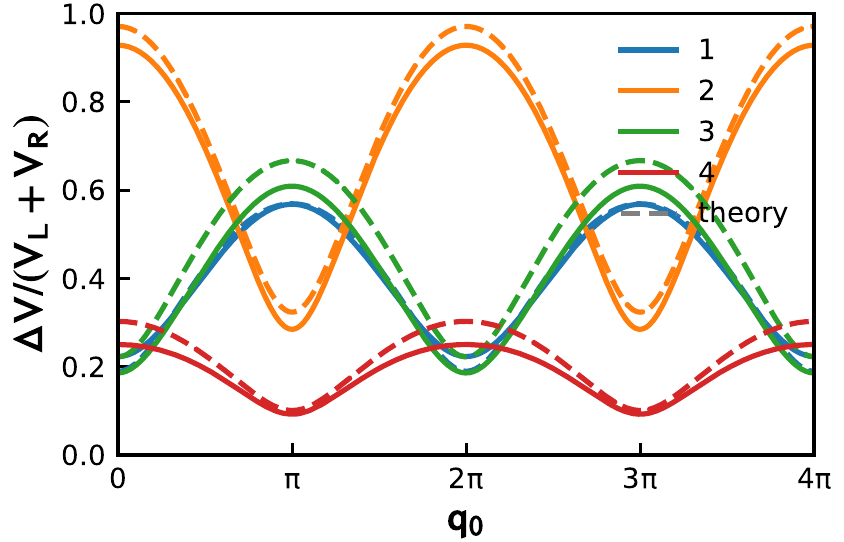}
  \includegraphics[width=0.24\textwidth]{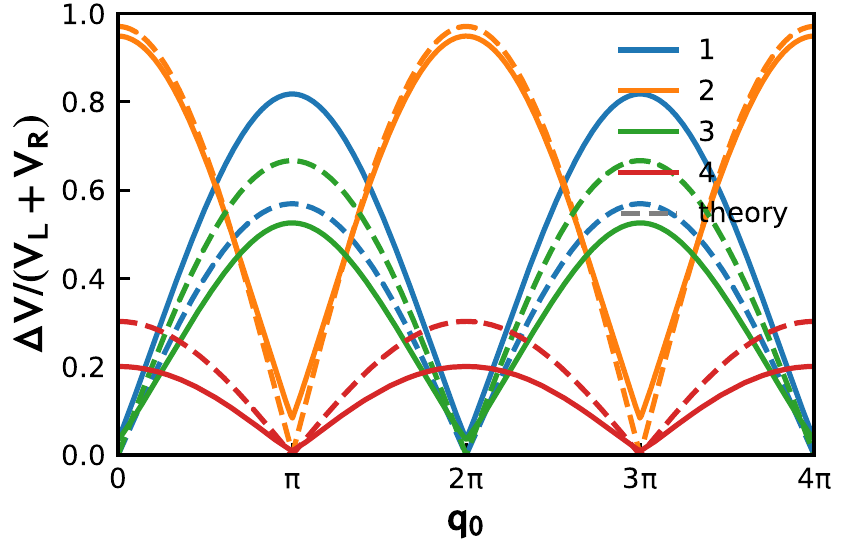}
  \caption{Strong coupling limit. The width of the current plateaux at $\omega=2\omega_0$. The Figures a, b, d are for symmetric setup. For Figures a, b, c $\tilde{C}=0.05$. The high-frequency Bessel approximation (Eqs. \ref{eq:stepsBessel},\ref{eq:2}) given by dashed lines. a. Width of several integer plateaus in symmetric case versus the driving amplitude at $q_0=\pi/2$. The curves are close to the high-frequency Bessel approximation. b. The widths versus $q_0$ for the constant amplitude $A=2\pi$. The destructive interference in the symmetric setup manifests as zero width and a cusp at the values $q_0=0, \pi$ for odd and even plateaux respectively. c. Same as in b. for a slightly asymmetric setup, $V_R=2V_L$. d. Same as in b., the capacitance is increased to $\tilde{C}=1$.} 
  \label{fig:saphiro-strong}
\end{figure*}

\begin{figure}[ht]
\label{fig:frequencydependence}
  \centering
  \includegraphics[width=\columnwidth]{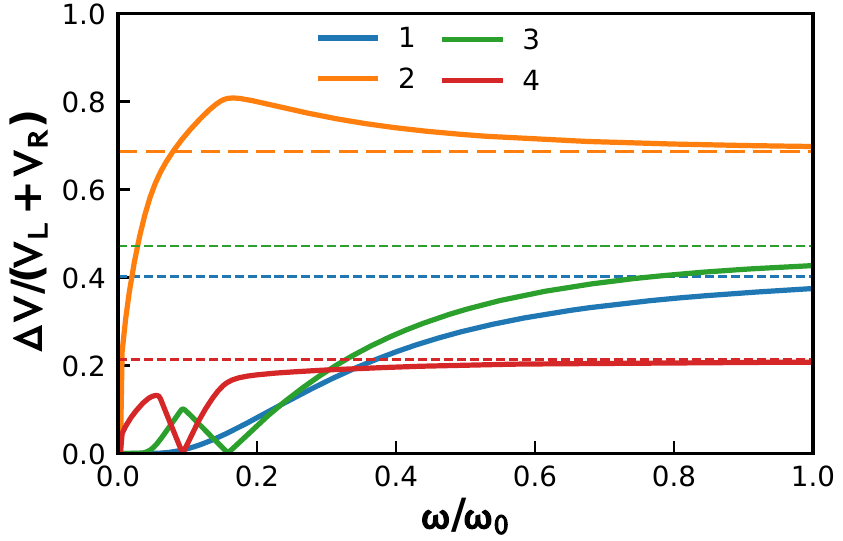}
  \caption{The frequency dependence of the plateau width for several integer platueax(see the labels). Strong coupling regime $\tilde{C}=0.2$, $q_0=\pi/2$, symmetric setup, the gate charge amplitude $A=2\pi$. The high-frequency limits (Eq.\ref{eq:stepsBessel},\ref{eq:2}) are given by dashed lines. The high-frequency approximation given (formula \ref{eq:stepsBessel}) is given by dashed lines. 
  We see that the limit is achieved at $\omega\approx \omega_0$ for all pleateaux. The width is  the high-frequency approximation is valid for all steps. The width decreases to zero at lower frequencies, with some non-monotonous Bessel-like dependence.}
\end{figure}

Let us derive an analytical expression for the width of the plateaux in the limit of $k\omega \gg \omega)$. It follows from Eq. \ref{eq:gate} that the limit $C\to 0$ $q_R=q_L+q_g(t)$. Let us separate $q_g$ into time-independent and oscillating part, $q_g(t) = q_0 + \tilde{q}_g(t)$.
Let us introduce a convenient variable
\begin{equation}
q = r_L q_L + r_R q_R + q_0(r_R -r_L - 1/2)
\end{equation} 
where $r^{R,L} = R_{R,L}/R_\Sigma$, 
Summing up Eqs. \ref{eq:junctions}, we obtain a convenient equation for this variable,
\begin{equation}
\label{eq:main}
\frac{e}{\pi} R_\Sigma \dot{q} + {\rm Im}\left[ e^{i q} Am(t)\right]=V
\end{equation} 
where the effective phase-slip amplitude $Am(t)$ reads
\begin{equation}
V_L e^{-i q_0/2} e^{-i\tilde{q}^R(t)} + V_R e^{i q_0/2} e^{i \tilde{q}^L(t)};  
\end{equation}
$\tilde{q}^{R,L}_g \equiv r_{R,L}  \tilde{q}_g$.

We concentrate on $V \gg (V_L+V_R)$ and $\pi V/eR_\Sigma$ close to $k\omega$. We will search for the solution in the form (see e.g. \cite{Likharev})
\begin{equation}
\label{eq:qhigh}
q = k \omega t + \psi(t)
\end{equation}
assuming the separation of time scales (see e.g. \cite{Likharev}), that is, the phase $\psi$ to change slowly on the scale of $\omega$. Substituting Eq. \ref{eq:qhigh} to Eq. \ref{eq:main} and averaging over the short time scale, we obtain an evolution equation for this slow phase
\begin{equation}
\label{eq:slow}
\frac{e}{\pi} R_\Sigma \dot{\psi} + {\rm Im}\left[ e^{i \psi} \overline{Am}_k\right] = \delta V,
\end{equation}
$\delta V = V - (e/\pi) R_\Sigma k \omega$, and $\overline{Am}_k$ is the result of averaging the amplitude over the oscillation period,
\begin{equation}
\overline{Am}_k = \int_0^{2\pi/\omega} dt Am(t) e^{i k\omega t}. 
\end{equation}
The Eq. \ref{eq:slow} has stationary solution for $\psi$ provided $|\delta V| < |\overline{Am}_k|$. This implies that the frequency of the oscillations in this voltage interval does not change being locked to $k \omega$. 
The width of the plateau is thus given by
\begin{multline}
\label{eq:stepsBessel}
\Delta V_k = 2|\overline{Am}_k| = \\
 2|V_L e^{-i q_0/2} J_k(A_R)  +V_R e^{i q_0/2} J_k(-A_L)|
\end{multline}
where in the last equation we have specified to the harmonic gate voltage signal,$ A_{R,L} \equiv r^{R,L} A$, $J_k$ being the Bessel function of the order $k$.
For symmetric case $R_R=R_L$, $V_L=V_R = V_\Sigma/2$, this becomes
\begin{align}
\Delta V_k = 2 V_\Sigma J_k(A/2) \begin{cases} |\cos(q_0/2)| , & \mbox{if } k\mbox{ is even} \\ |\sin(q_0/2)|, & \mbox{if } k\mbox{ is odd} \end{cases} \label{eq:2}
\end{align}
We thus have selection rules in this case: no odd plateaux at $q_0$, no even plateaux at $q_0 = \pi$.

Although the expression (\ref{eq:stepsBessel}) is formally valid only in the limit $k\omega \gg \omega_0$, we find numerically that it gives qualitatively good estimations for all $\omega \simeq \omega_0$. We extract the plateau widths from numerical data finding the voltages at which the relative deviation of the current from the quantized value amounts to $10^{-3}$ and associating those with the endpoints of the locking interval. 

We present the numerical results for $\omega =2 \omega_0$ in Fig. \ref{fig:saphiro-strong} in comparison with Eqs. \ref{eq:2}, \ref{eq:stepsBessel} . Fig. \ref{fig:saphiro-strong} a gives the widths versus the driving amplitude $A$ for the symmetric setup at  $q_0=\pi/2$ so that the plateaux of both parities are developed. As we see, the actual widths coincide with Bessel function prediction with the accuracy of several per cent. In Fig. \ref{fig:saphiro-strong} b, we plot the widths versus $q_0$ at constant driving amplitude. We observe the selection rules mentioned: the width drops to 0 with the plotting accuracy for $q=0$ and odd plateaus, and for $q_0=\pi$ and even plateaus, manifesting the descructive interference of the phase slips. The curves make a cusp at this values of $q$. As we see in Fig. \ref{fig:saphiro-strong} c, the curves are smooth and do not reach 0 if we depart from the symmetric case ($V_L/V_R =2$ in this Figure). 
It is interesting to note that if we decrease the capacitance moving towards the intermediate coupling regime,  ($\tilde{C} =1.0$ for Fig. \ref{fig:saphiro-strong} d), the cusps in the symmetric case do not dissapear but visibly depart from 0. Thus the small decoupling of the phase slips supresses the destructive interference. This is consistent with the results for the Coulomb blockade threshold voltage described in the previous Section.

The analytical expressions for widths given by Eqs. \ref{eq:stepsBessel},\ref{eq:2} do not depend on frequency. This should not be valid for sufficienly small frequency. Indeed, we see that the plateaux disappear in the limit of $\omega \ll \omega_0$. in the limit of small frequency.(Fig. \ref{fig:frequencydependence}) We also see that the high-frequency limit is achieved already at $\omega \approx \omega_0$ for all plateaux and the frequency dependence is generally non-monotonous resembling the Bessel-like dependence on the amplitude.

\section{Fractional plateaux in intermediate coupling regime}\label{sec:fractional}

\begin{figure}[ht]
  \centering
  \includegraphics[width=\columnwidth]{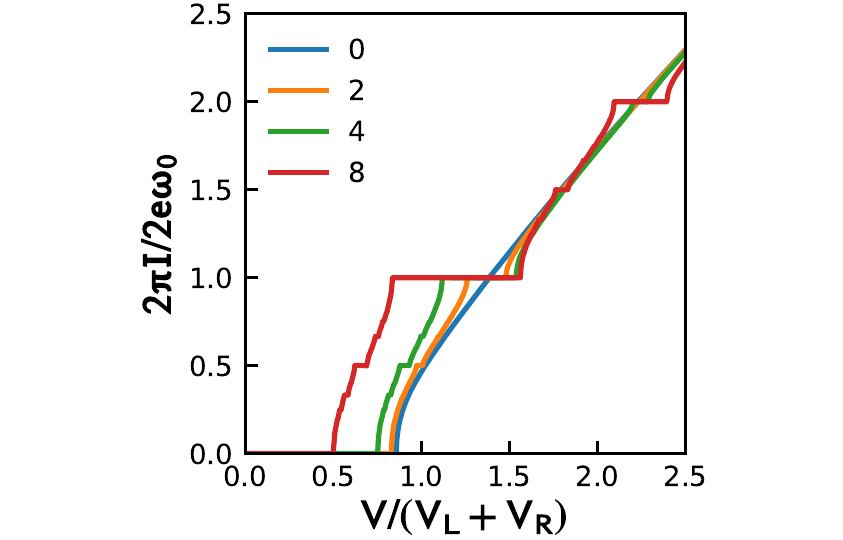}  
 \caption{Indermediate coupling regime $\tilde C = 5$. Symmetric setup. I-V characteristics  for $q_0=\pi/2$ and several different driving amplitudes $A$ given in the labels. In addition to the integer plateaux, we see the emergence of smaller fractional plateaux at $I_{MN} =  (e/\pi) \omega \frac{M}{N}$.}\label{fig:inter-iv}
\end{figure}

In this Section, we will discuss the intermediate coupling regime $\tilde{C} \simeq 1$. Typical I-V characteristics with driving are presented in Fig. \ref{fig:inter-iv}. A striking difference from the strong coupling regime is the appearance of smaller current plateaux at integer fractions of $e\omega/\pi$. In principle, we expect a plateau at any rational fraction $I_{MN} =  (e \omega/\pi) {M}/{N}$, $M$, $N$ being integer, so that the actual I-V characteristics resemble Cantor function and is an example of devil's staircase \cite{Bak1986}. 
In practice, the plateau widths become exponentially small upon increasing $N$ so only several fractional plateaux are visible. In our numerics, we were able to detect the features up to $N=7$.  The fractional plateaux are best visible for $\tilde{C} \simeq 5-10$ and will gradually disappear upon further increase of capacitance, see the next Section.

To understand the emergence of fractional plateaux analytically, we will develop a perturbation theory in terms of small $\tilde C$. To simplify the derivations, we resort to the fully symmetric setup ($R_L=R_R=R_\Sigma /2$, $V_L=V_R=V_\Sigma/2$).

It follows from Eqs. \ref{eq:gate}, \ref{eq:junctions} that the first-order correction in $\tilde{C}$ to Eq. \ref{eq:main} can be presented as a small change of the gate charge 
\begin{multline}
    q_g(t) \to q_g(t)  - \frac{\pi C}{2e} \left(
    \frac{e}{2 \pi} R_\Sigma \dot q_g(t)\right. \\
	\left.+ V_\Sigma \sin \frac{q_g(t)}{2} \cos q
    \right) 
\end{multline}
The first term in the addition is an insignificant modification of the signal while the second term brings higher harmonics of $e^{iq}$ into the Eq. \ref{eq:main} that becomes

\begin{figure}[ht]
  \centering
  \includegraphics[width=\columnwidth]{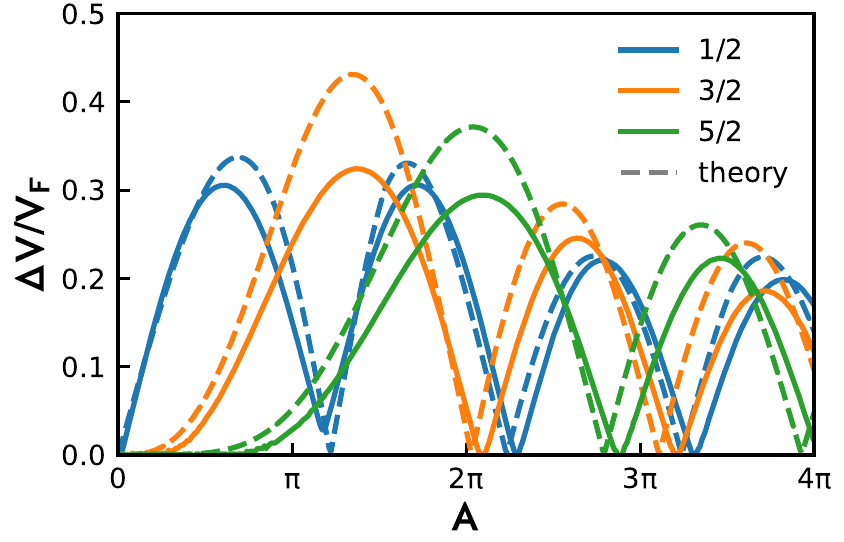}
  \caption{The widths of the half-integer plateaux 1/2, 3/2, and 5/2 versus the driving amplitude. Symmetric setup, $\omega = 2 \omega_0$, $q_0 = \pi/2$, $\tilde C = 0.2$. The numerical results (solid lines) are in good correspondence with the semi-analytical prediction (Eq.\ref{eq:heuristic}, dashed lines ) } \label{fig:inter-step}
\end{figure}

  \begin{multline}
  \label{eq:withVF}
    \frac{e}{ \pi} R_\Sigma \dot q = V - V_\Sigma \cos \frac{q_g(t)}{2} \sin q  \\
	- V_F \sin^2 \frac{q_g(t)}{2} \sin 2q. 
	\end{multline}
Here $V_F \equiv{\pi C V_\Sigma^2}/{8e}$, $V_F \ll V_\Sigma$, and the third term represents a relevant correction responsible for emergence of the half-integer plateux. A similar equation has been studied in the context of fractional Shapiro steps.\cite{Romeo2004} Further orders in $\tilde{C}$ would provide the terms $\propto e^{i N q}$ with $N >2$ that will account for the plateaux at higher fractions.

We analyze this in the limit of high frequencies searching for the solutions in the form (c.f. Eq.\ref{eq:qhigh}) 
\begin{equation}
\label{eq:qhighhalf}
q = (k + 1/2) \omega + \psi(t)
\end{equation}
$\psi(t)$ being the slow-varying phase. We can neglect the second term in Eq. \ref{eq:withVF} since it averages out over the period. The resulting equation for the phase is contributed by the third term,
\begin{equation}
\label{eq:fortwopsi}
\frac{e}{\pi} R_\Sigma \dot{\psi} + {\rm Im}\left[ e^{2 i \psi} \overline{Am}^{(2)}_k\right] = \delta V
\end{equation}
where 
\begin{equation}
\overline{Am}^{(2)}_k \equiv \frac{V_F \omega}{2\pi} \int_0^{\frac{2\pi}{\omega}} dt \sin^2 \frac{q_g(t)}{2} e^{i (2k + 1)\omega t}.
\end{equation}
This gives the width of the half-integer plateau
\begin{equation}
\label{eq:half-integer-bessel}
\Delta V_{k+1/2} = V_F |J_{2k+1}(A)| |\sin q_0| 
\end{equation}
which is parametrically smaller than $V_\Sigma$ in the limit $\tilde{C} \to 0$ and is of the order of $V_F$ in the intermediate coupling regime.

This gives an accurate prediction at high frequencies $\omega \gg \omega_0$. To investigate the half-integer phase-slips at lower frequencies, we build up a more complex perturbation theory in $\tilde{C}$. This relies on a heuristic assumption, however, it accurately and adequately describes complex numerical data (see Fig. \ref{fig:inter-step}).

To start with the low-frequency perturbation theory, we would need to solve the unperturbed equation 
\begin{align}
  \label{eq:withoutVF}
    \frac{e}{ \pi} R_\Sigma \dot q = V - V_\Sigma \cos \frac{q_g(t)}{2} \sin q
\end{align}
We cannot find an explicit analytical solution. Instead, we use a solution $q_{\star}(t)$ of an {\it autonomous} equation obtained by the averaging of $\cos ({q_g(t)}/{2})$ over the period, 
\begin{align}
  \label{eq:qstar}
    \frac{e}{ \pi} R_\Sigma \dot q_\star = V - V_\Sigma \bar{f} \sin q_\star, 
\end{align}
$\bar{f} = J_0(A/2) \cos(q_0/2)$.This equation can eventually be solved \cite{Likharev} and we write the solution in the form
\begin{align}
  \label{eq:starsolution}
    \frac{dt}{d q_\star} = \frac{V - V_\Sigma\cos(\Omega t +\psi)}{\Omega^2 \pi R/e}
\end{align}
where the frequency of autonomous oscillations 
\begin{equation}
\Omega= \pi \frac{\sqrt{V^2 -V^2_\Sigma \bar{f}^2}}{e R}
\end{equation}.
We are interested in $\Omega$ close to $M \omega/2$, $M$ being odd. We substitute the expression \cref{eq:starsolution} to Eq.\ref{eq:withVF} and derive an equation for the slow-varying phase that is similar to Eq. \ref{eq:fortwopsi}. This finally gives a semi-analytical expression for the plateau width: 

\begin{widetext}
  \begin{multline}
  \label{eq:heuristic}
    \frac{\Delta V_{M/2}}{V_F} = \frac{(\pi R/e) M^2}{V \omega_0\bar f^2} (\max_\phi - \min_\phi) \left\{
      M \frac{\omega}{\omega_0} \sin \frac{q_0}{2}  \sum_{kM \in odd} J_{kM}(A)\left(\frac{\bar f}{y}  \right)^{2k} \sin (2 k \phi) \right. \\
    \left.
      - M\frac{\omega}{\omega_0} \cos \frac{q_0}{2}  \sum_{kM \in even} J_{kM}(A)\left(\frac{\bar f}{y}  \right)^{2k} \cos (2 k \phi)
      - \frac{\kappa}{2} (1 - J_0(A) \cos \frac{q_0}{2}) J_M(A) \cos \frac{q_0}{4} \sin \phi
    \right\}   
  \end{multline} 
\end{widetext}

  where 
\begin{align}
    y &= \sqrt{V/V_\Sigma - \bar{f}^2} +V/V_\Sigma;\; w = \Omega/\omega_0; \; v = V/V_\Sigma\nonumber
    \\
    \kappa &= \bar f \left( \frac{v^2}{w^3} - 4 \frac{w}{y^2} \right)
    - \bar f^3 \left( \frac{2v}{3 y^2 w} + \frac{v(w + y)}{w^3 y^2} \right) \nonumber
  \end{align}
This coincides with Eq. \ref{eq:half-integer-bessel} in the limit of high frequencies and   
diverges at $\omega \to 0$ invalidating the perturbation theory in this limit.

  \begin{figure}[ht]
    \centering
  \includegraphics[width=\columnwidth]{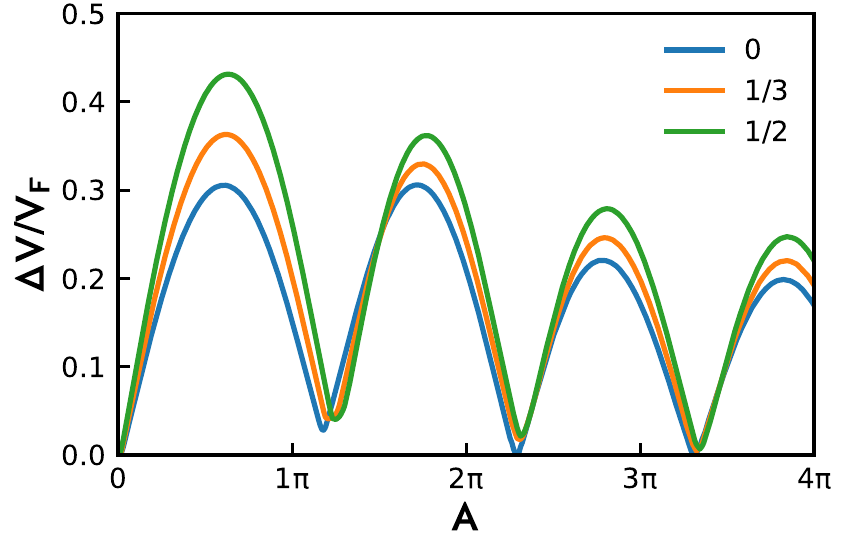}    
  \includegraphics[width=\columnwidth]{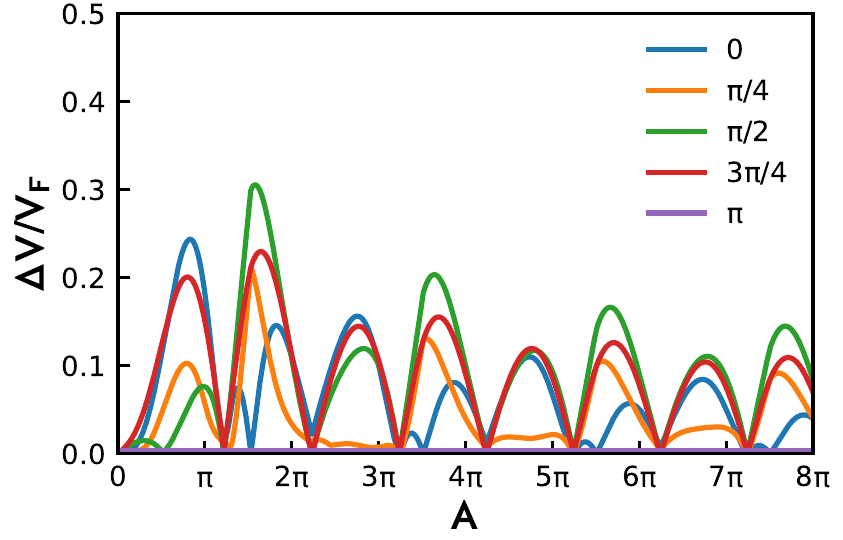}    
  \caption{The widths of 1/2 plateau in dependence on asymmetry of the junctions and $q_0$, $\omega = 2\omega_0$, $\tilde C=0.2$. Upper pane: The width at $q_0 = \pi/2$ versus the driving amplitude for several values of the asymmetry parameter $(V_L - V_R)/(V_L + V_R)$ shown in the labels. Lower pane: The width at $\omega = \omega_0$ versus the driving amplitude for several values of $q_0$ given in the labels. The setup is symmetric. We observe a rather complex dependence coming from the contributions of various Bessel functions. The data are obtained from Eq. \ref{eq:heuristic}. No plateaux persist at $q_0 =\pi$. }
  \label{fig:fancy}
  \end{figure}

  In Fig. \ref{fig:inter-step}, we plot the widths of the half-integer plateux for a small capacitance $\tilde C =0.1$ so we can compare them with the semi-analytical prediction obtained, Eq.\ref{eq:heuristic}. The deviations such as the shift of position of the cusps, the finite width at the cusp, and the heights of maxima, arise from  the higher-order terms in $\tilde{C}$.
  
  More details about the width of the half-integer plateaux illustrating the effects of junction asymmetry and $q_0$ are presented in Fig.\ref{fig:fancy}.

\section{Plateaux in the Weak coupling regime}\label{sec:weak-coupling-regime}

\begin{figure}[ht]
  \centering
  \includegraphics[width=\columnwidth]{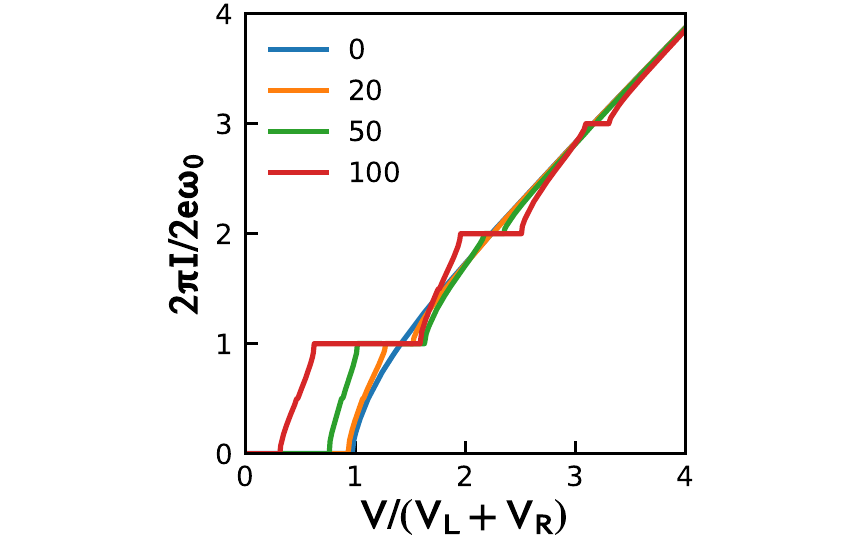}  
  \caption{Plateaux in the weak coupling regime. Symmetric setup, $\tilde C = 50$,$\omega=\omega_0$, $q_0=\pi/2$. The fractional plateaux are hardly visible for such values of $\tilde{C}$ while the integer plateaux remain pronounced. The labels give the induced charge modulation amplitude $A$ as in the previous Sections. We see however the change of the scaling: the plateaux become of the order of $V_{L}+V_{R}$ at $A \simeq \tilde{C}^{-1}$, that is, at the gate voltage modulation amplitude $\tilde{V}_g \simeq V_{L}+V_{R}$.  In distinction from the strong coupling regime, the plateau width at a given amplitude is decreasing upon increasing frequency or voltage, $\Delta V \propto \omega^{-1}$. } \label{fig:weak}
\end{figure}

The weak coupling regime occurs at sufficiently big capacitances, $\tilde C \gg 1$. In this limit, the capacitor can be regarded as a voltage source that completely decouples the junctions with respect to a.c. voltage. Of course, the d.c. coupling still persist so the same  d.c. current flows through the junctions, but all interference effects characterized by $q_0$ dependence eventually disappear in this limit as well as the fractional current plateaux. Perhaps unexpectedly, the absence of $q_0$ dependence does not suppress the synchronization of Bloch oscillations by the gate voltage: we see in Fig. \ref{fig:weak} well-developed integer current plateaux.

\begin{figure*}[ht]
  \centering
  \includegraphics[width=\columnwidth]{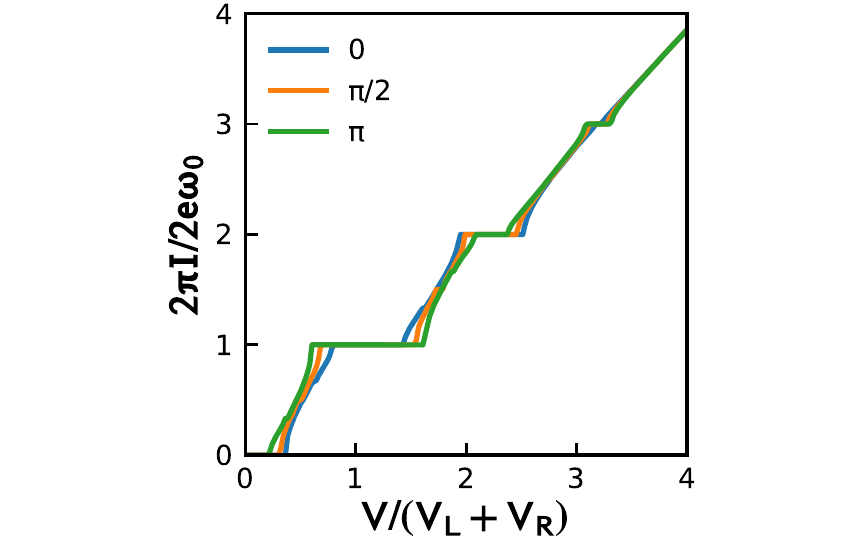}
  \includegraphics[width=\columnwidth]{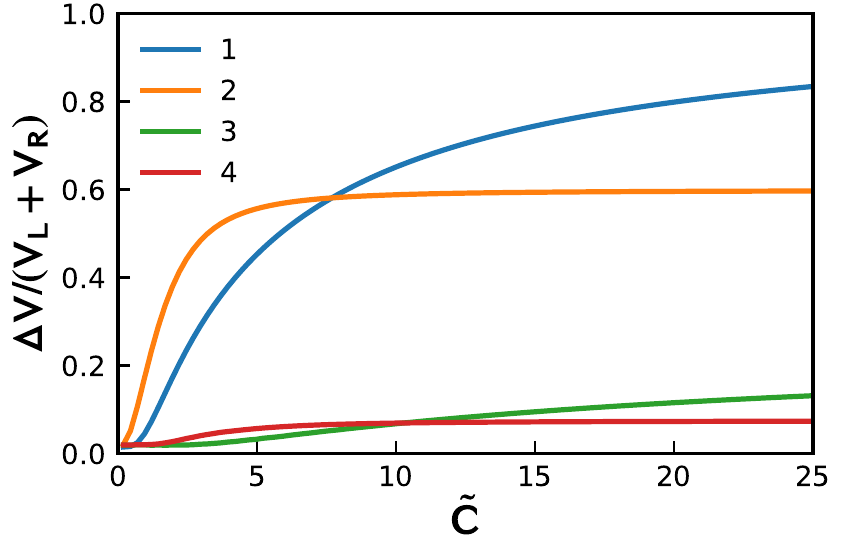}
  \caption{Approachig the weak coupling regime. Symmetric setup. Left:  I-V characteristics at $\tilde C =10$, $A=20$ and several values of $q_0$ given in the labels compared with the I-V characteristic of the single junction with $\tilde{V}_g$ added to bias voltage (Eq. \ref{eq:single}). All these curves should coincide in the limit $\tilde{C} \to \infty$. For finite $\tilde{C}$, we see some residual deviations modulated by $q_0$ and even small fractional plateaux. Right: The plateau width (in units of $V_L +V_R$ of several integer plateaux at fixed a.c amplitude $\tilde{V}_g$ (same as that on the left) versus the capacitance $\tilde{C}$. The saturation indicates the weak coupling regime, and occurs at different $\tilde{C}$ for different plateaux.  } \label{fig:weak2}
\end{figure*}

To obtain the dynamical equations in this regime, we notice that the big capacitors effectively short-cuts $V_N(t)$ except zero frequency. So in this limit 
$V_N(t) =\tilde{V}_g(t) + \bar{V}$, $\bar{V}$ not depending on time, $\tilde{V}_g(t)$ has no d.c. component. With this, Eqs. \ref{eq:junctions} become
:
\begin{align}
   \bar{V} + \tilde{V}_g(t)&= V_R \sin q_R + \frac{e}{\pi} R_R \dot q_R \\
  V - \bar{V} -\tilde{V}_g(t)&= V_L \sin q_L + \frac{e}{\pi} R_L \dot q_L\label{eq:12}
\end{align}
So the equations for left and right junctions separate, each is for a single junction biased by a d.c and a.c. voltage. The only coupling is provided by $\bar{V}$ that is determined from the d.c. current conservation:
\begin{equation}
\bar{I}_L(V-\bar{V}) = \bar{I}_R(\bar{V}).
\end{equation}
 In general situation, the solution $\bar{V}$ of this equation is  formally ambiguous right at a current plateau, since at the plateau the currents do not depend on voltages. This ambiguity, however, is readily resolved if voltage noise is taken into account. As we discuss in the next Section, this results in finite differential conductance at the plateaux and unambiguous solution. In any case, the total width of a plateau is just the sum of the widths for constituent junctions,
\begin{equation}
\Delta V = \Delta V^L + \Delta V^R.
\end{equation}
For a symmetric setup, this gives $\bar{V} = V/2$ and the $I-V$ characteristic is the same as for a single junction at half voltage and $\tilde{V}_g$ added to the bias voltage,
\begin{equation}
\label{eq:single}
I(V) = I_{{\rm single}}(V/2).
\end{equation}

 Let us obtain the analytical prediction for the plateau width in the limit of big frequency. We concentrate on the left junction and chose a harmonic drive:
 \begin{align}
  \frac{e}{\pi} R_L \dot q = V - \bar{V} - V_L \sin q - \tilde{V}_g \sin (\omega t)
\end{align}
Near the $k$-th integer plateau, we neglect $V_L$ and search the solution in the form 
\begin{equation}
q(t) = k\omega + \frac{\pi \tilde{V}_g}{e R_L} \cos \omega t + \psi(t).
\end{equation}
The resulting equation for the slow phase is very similar to Eq. \ref{eq:slow}:
\begin{equation}
\label{eq:slow2}
\frac{e}{\pi} R_\Sigma \dot{\psi} + {\rm Im}\left[ e^{i \psi} \overline{Am}_k\right] = \delta V,
\end{equation}
where 
\begin{equation}
\overline{Am}_k = \frac{V_L \omega}{2\pi} \int_0^{\frac{2\pi}{\omega}} dt \exp\left(i k \omega t + i \frac{\pi \tilde{V}_g}{e R_L} \cos \omega t\right) 
\end {equation}
This gives the plateau width for the left junction
\begin{equation}
\Delta V^L = 2 V_L \left|J_k\left(\frac{\pi \tilde{V}_g}{e \omega R_L}\right)\right|
\end{equation}
and the overall width 
\begin{equation}
\label{eq:stepsBessel2}
\Delta V^L = 2 V_L \left|J_k\left(\frac{\tilde{V}_g}{e \omega R_L}\right)\right| + 2 V_R \left|J_k\left(\frac{\tilde{V}_g}{e \omega R_R}\right)\right|
\end{equation}
This result is somewhat similar to that for the strong coupling regime (Eq. \ref{eq:stepsBessel2}): the maximum width is restricted by $V_L + V_R$ and exhibit the Bessel-like dependence on the driving amplitude. We note however the disappearance of interference and renormalization of the arguments in the Bessel functions.
The arguments are inversely proportional to $\omega$, this reduces the widths upon increasing $\omega$. However, much smaller amplitudes $\tilde{V}_g$ are required for the argument of the Bessel functions to be of the order of 1: if $\omega \simeq \omega_0$, 
$\tilde{V}_g \simeq V_{L,R}$ as opposed to $\tilde {V}_g \simeq V_{L,R} /\tilde{C} \gg V_{L,R}$ in the strong coupling regime.

The weak coupling limit described corresponds to $\tilde{C} \to \infty$, and at any finite capacitance up to $\tilde{C} \simeq 10^2$ there are still noticeable deviations. We illustrate this in Fig. \ref{fig:weak2}. In the left pane, we plot $I-V$ characteristics at different $q_0$ for $\tilde{C} =10$. The $q_0$ dependence should be absent in the weak coupling limit, and the curves should collapse on the single-junction $I-V$ characteristic. We see, however, sizeable deviations and even small fractional plateaux.
In the right pane, we plot the widths of the integer plateaux at fixed gate voltage modulation amplitude $\tilde{V}_g$ (since $A \propto \tilde{C}$, the widths vanish at $\tilde{C} \to 0$. The widths should saturate in the weak coupling limit. We see that the saturation is slow and also different for different plateaux.

\begin{figure*}[!ht]
  \centering
  \includegraphics[width=\columnwidth]{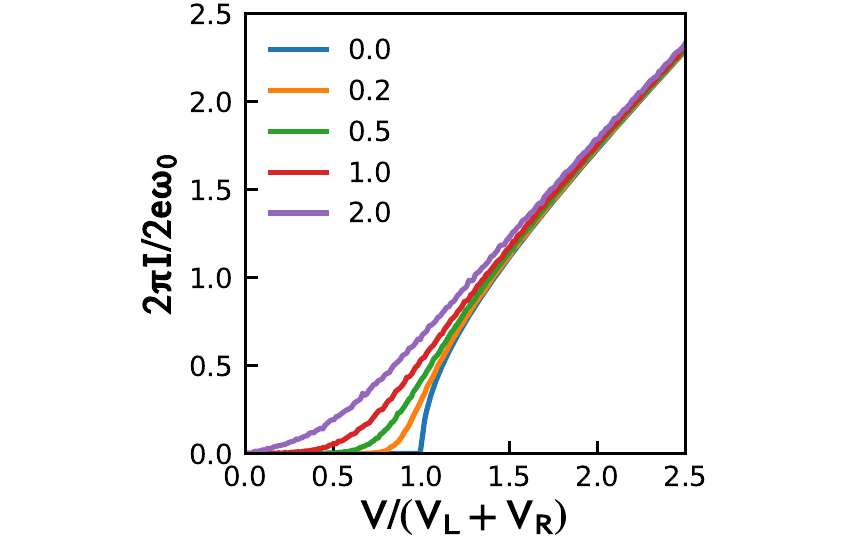}
  \includegraphics[width=\columnwidth]{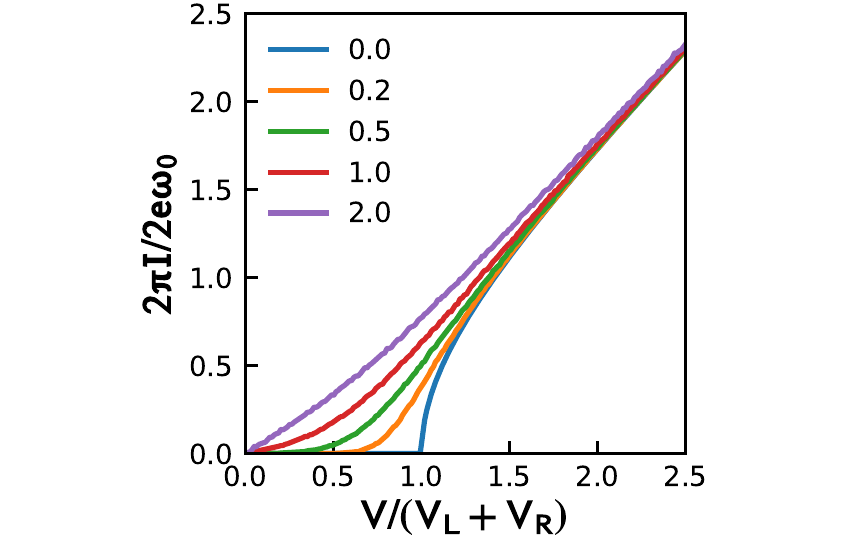}
   \label{fig:columb-noise}
  \caption{Smoothing of the Coulomb blockade feature by finite temperature. I-V characteristics for several dimensionless temperatures $\tilde{T} = 8 \pi k_B T/e V_\Sigma$. Symmetric setup, $q_0=0$. Left: $\tilde{C} =0.1$, strong coupling regime. Right: $\tilde{C} =50$, weak coupling regime. Upon increasing the temperature, we see first the rounding of the sharp feature at the threshold, them finite conductance at zero voltage and finally a linear I-V characteristic. The same degree of smoothening in the weak coupling regime occurs at approximately half of the temperature at which it occurs in the strong coupling regime. } 
\end{figure*}

\section{Finite temperature effects} \label{sec:finite-temp-effects}

\begin{figure*}[ht]
  \centering
  \includegraphics[width=\columnwidth]{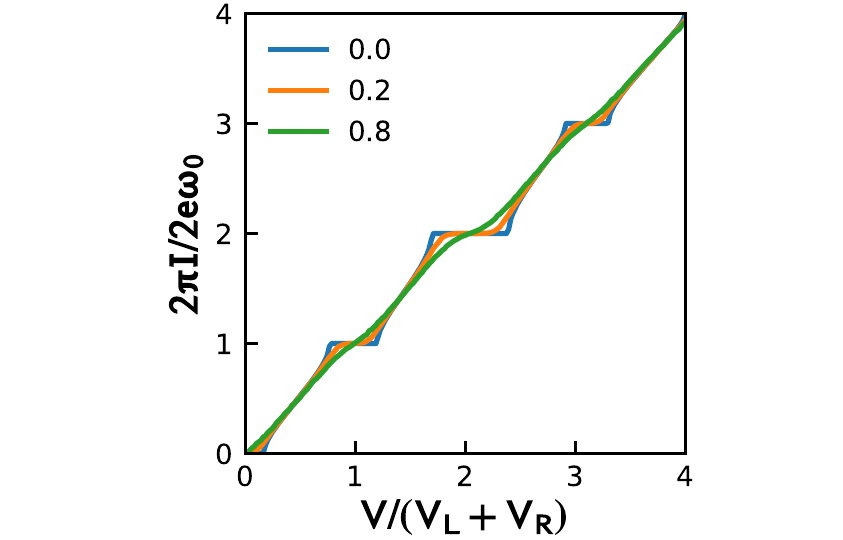}
  \includegraphics[width=\columnwidth]{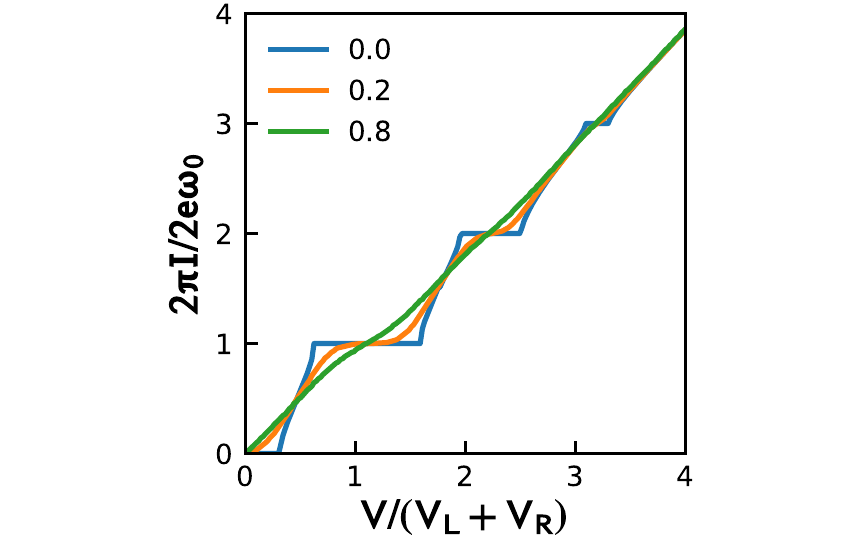}
  \caption{Smoothing of current plateaux at finite temperature. I-V characteristics of symmetric setup under gate voltage modulation with $\omega = \omega_0$ at dimensionless temperatures $\tilde{T}$ shown in the labels. Left: strong coupling regime, $\tilde C = 0.2$, $q_0=\pi/2$, the induced charge modulation $A = 6$. Left: weak coupling regime, $\tilde C = 50$, the voltage gate modulation $A=100$ with $q_0 = \pi/2$.  
  At the same temperature, the degree of smoothing is bigger for the plateaux of smaller width. The same degree of smoothing in the weak coupling regime occurs at twice smaller temperature as compared with the strong coupling regime.} \label{fig:saphiro-noise}
\end{figure*}

In the previous Sections, we neglected the finite temperature that in our semiclassical model is manifested as a white voltage noise. This permitted us to concentrate on ideal synchronization, sharp Coulomb blockade and plateau features. In this Section, we investigate how the synchronization is gradually destroyed by noise. This is manifested as gradual smoothing eventual disappearance of blockade and plateau features. This study is especially relevant in the setups including large resistors in view of the dissipation and resulting overheating of the resistors. We numerically solve the Eqs. \ref{eq:1}at large time intervals to obtain the I-V characteristics with and without modulation. The main goal of this study is to come up with approximate but practical estimations of the temperatures at which the plateaux are still observable. With this, we can also draw semi-quantitative predictions beyond semi-classics using the correspondence between the thermal and quantum noise at a plateau  developed by modulation with the frequency $\omega$  \cite{Hriscu2013} : $k_B T = \hbar \omega$.

We present first the finite temperature effect on the Coulomb blockade feature in the absence of the a.c. modulation (Fig.\ref{fig:columb-noise}). The temperature scale at which Coulomb blockade deteriorates should correspond to the Coulomb energy $e V_\Sigma$. We incorporate this introducing a dimensionless temperature $\tilde{T} \equiv 8 \pi k_B T/e V_\Sigma$. The smoothing of the Coulomb blockade for our setup is qualitatively similar to the smoothing of I-V characteristics in common Coulomb blockade: at low temperatures, the sharp corner at the threshold is rounded at the scale of this low temperature while the differential conductance remains strongly suppressed; at medium temperatures, the low-voltage conductance becomes a fraction of $R^{-1}$; at higher temperatures, the Coulomb blockade feature disappears and the I-V characteristics is almost linear. In Fig. \ref{fig:columb-noise}, we compare the results for strong and weak coupling regimes choosing $q_0=0$ where the Coulomb blockade thresholds are maximized and equal in both limits. We see that for the symmetric setup the same degree of smoothing in the weak coupling regime requires twice smaller temperature as compared to the strong coupling regime. This is explained by the fact that the height of the Coulomb barrier is twice smaller in the weak coupling regime. Indeed, since the junctions are uncoupled, the Coulomb barriers are determined by $e V_{L,R}$ as compared to $e(V_L + V_R)$ in the strong coupling regime.

The smoothing of the well-developed integer plateaux with $\Delta V \simeq V_\Sigma$ follows the same pattern (Fig. \cref{fig:saphiro-noise}).  Also here the same degree of smoothing occurs at approximately twice smaller temperature in the weak coupling regime. One can see it, for instance, for the second plateau that is slightly wider in the weak coupling regime but is more smoothed at the same temperature. We also see that the smaller plateaux are smoother at the same temperature. This leads us to a simple scaling hypothesis: for each plateau, the degree of smoothing is defined by the temperature relative to the plateau width (c.f. \cite{Likharev}). To check the hypothesis, we need to choose a measure of smoothing.

\begin{figure}[ht]
  \centering
  \includegraphics[width=\columnwidth]{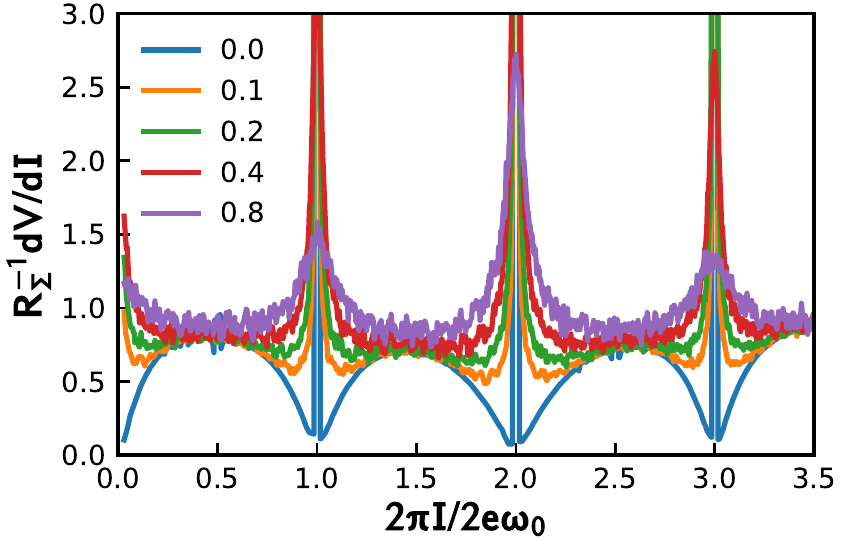}
  \caption{The differential resistance versus current. Symmetric setup, $\tilde{C} = 0.2$, a.c. voltage modulation at $\omega = \omega_0$, $A=6$, the dimensionless temperature $\tilde{T}$ is given in the labels. The resistance is  obtained numerically from I-V characteristics taken at discrete values of voltage with the current step $0.03 \frac{2 e \omega_0}{2 \pi}$. The noise in the data comes from the actual noise: To obtain I-V curves, we average the current over a finite time interval picking up its fluctuations. The smoothed plateaux are manifested as the peaks of differential resistance. The height of the peaks decreases with increasing temperature. We define the "width" of the plateau at finite temperature as the length of the voltage interval where the differential resistance exceeds the background differential resistance $\simeq R$ by at least a factor of 3.  } \label{fig:peaks}
\end{figure}

To do this, we note that a common experimental signature of imperfect plateaux are the peaks in differential resistance. Typical current dependences of the differential resistance for various temperatures are plotted in Fig. \ref{fig:peaks}. There, we see the peaks at the quantized values of the current that diverge at vanishing temperature, take finite value at finite temperatures and eventually merge with the background $\approx R_\Sigma$ at higher temperatures. We define the width of the plateaux at finite temperature as the length of the voltage interval where the differential conductance exceeds the background resistance at least  by a factor of $Q$, and choose $Q=3$. With this definition, the plateau widths becomes zero at some critical temperature where the peak differential resistance is thrice the background.  Despite the arbitrariness of this definition, it seems to be a reasonable practical compromise. To extract so-defined  width numerically, we change voltage in small steps near the quantized values of the current checking the differential conductance at each step.

The simple scaling hypothesis would imply that the temperature-dependent width in units of zero-temperature width is a universal function of temperature in units of the zero-temperature width,
\begin{equation}
\frac{\Delta V(T)}{\Delta V(0)} = f\left(\frac{k_B T}{e \Delta V(0)} \right)
\end{equation}
To check the hypothesis, we plot the evaluated widths of several plateaux in coordinates ${\Delta V(T)}/{\Delta V(0)}$, $\tilde{T} V_\Sigma/\Delta V(0)$. We plot the data both for strong and weak coupling regime, correcting the temperature by a factor of 2 in for the latter case. We see a good collapse of the data into a single curve despite significantly different widths of the plateaux. We concude that the width of a plateau is halved at $k_B T \approx 0.03 e \Delta V(0)$ and vanishes at 
\begin{equation}
k_B T_c \approx 0.06 e \Delta V(0).
\end{equation}  

\begin{figure}[ht]
  \centering
    \includegraphics[width=\columnwidth]{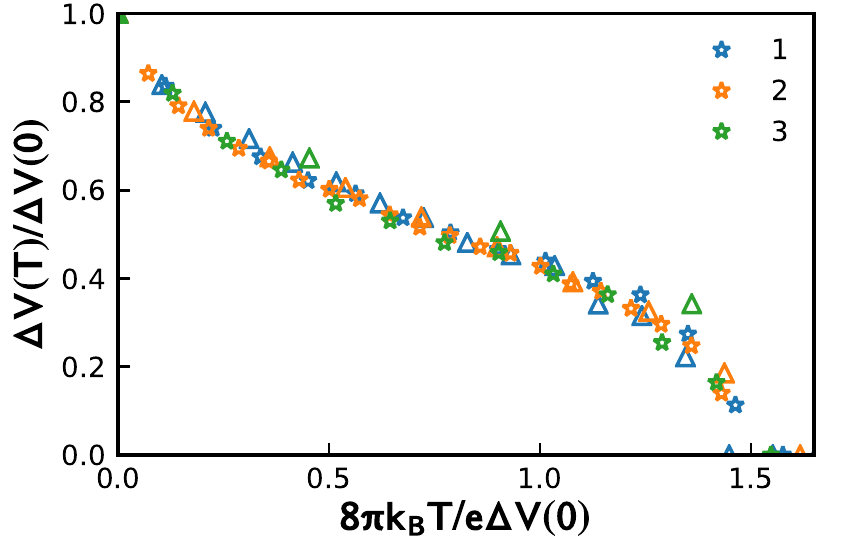}
    \caption{
      The scaling of the integer plateau smoothing with temperature. We check a simple scaling hypothesis
    $\Delta V(T)/\Delta V(0) = f(k_B T/e V(0))$ and see the data collapsing into the same universal curve for both weak and strong coupling regime and various plateaux. For the weak coupling regime, we correct the temperature by a factor of 2. The data for the strong coupling regime ($\tilde C = 0.2$, $A= 6$, $q_0 = \pi/2$) are plotted with stars, those for the weak coupling regime ($\tilde C = 50$, $A= 100$) are plotted with trianlge, the color of the star/triangle corresponds to first, second, third plateau as shown in the labels, $\omega = \omega_0$ for all situations. 
   } \label{fig:steps-T}
\end{figure}

The suggested scaling is not exact in any obvious limit, in fact, since we have to correct the weak coupling regime data, it would not work in between the regimes at $\tilde{C} \simeq 1$. Albeit it seems to work empirically.

To convert it into a quantum noise estimation, we substitute $k_B T_c = \hbar \omega$ and $\omega = \omega_0$. This gives a minimum value of the resistance $R_\Sigma$ at which the plateau is still observable,
\begin{equation}
R_c \approx 16 \frac{\pi \hbar}{e^2} \frac{ V_\Sigma}{\Delta V(0)}.
\end{equation}

\begin{figure}[ht]
  \centering
  \includegraphics[width=0.45\columnwidth]{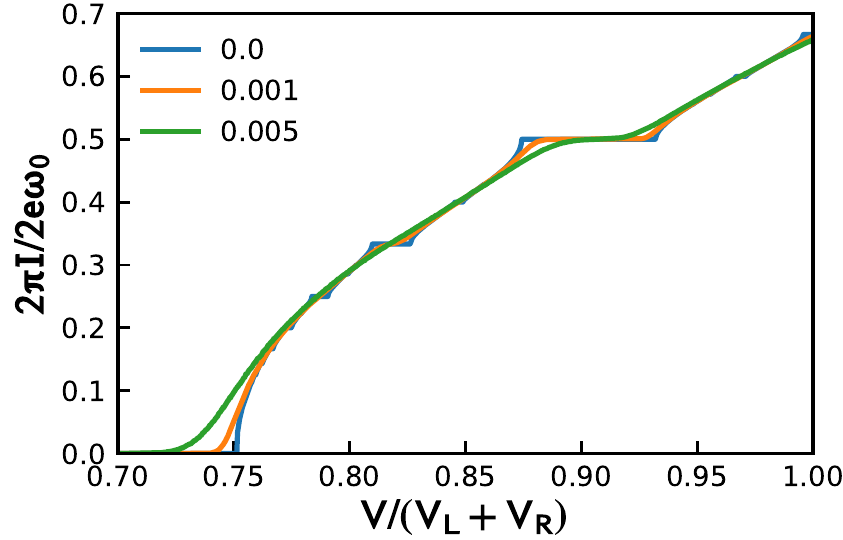}
  \includegraphics[width=0.45\columnwidth]{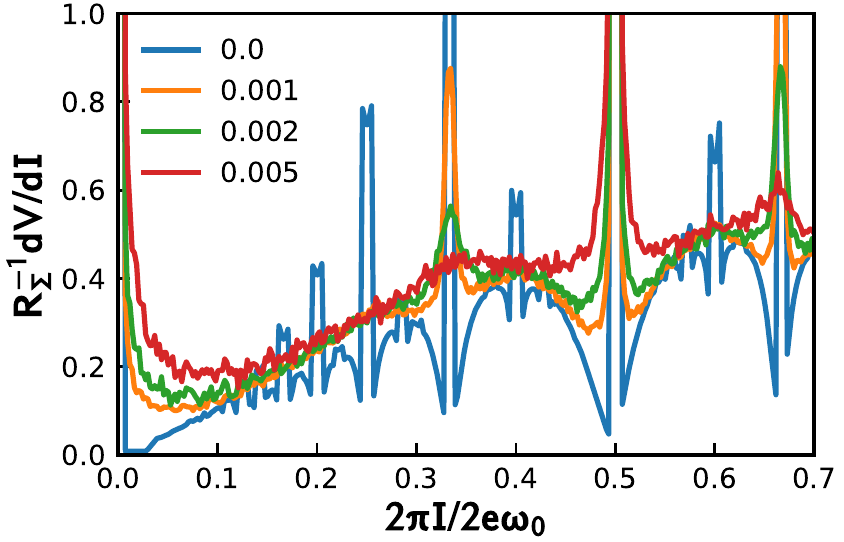}
  \includegraphics[width=\columnwidth]{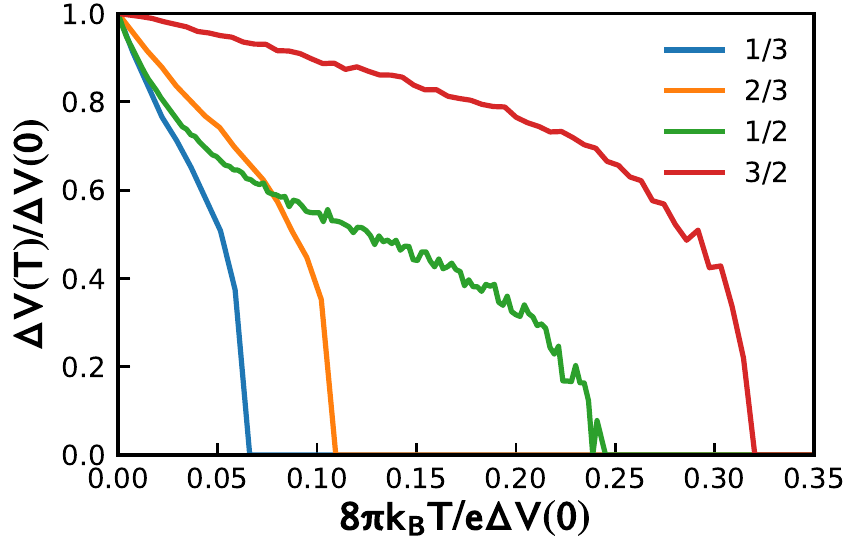}
  \caption{Failure of the simple scaling hypothesis for fractional plateaux. The data are for symmetric setup, $\omega = \omega_0$, $A= 4$, $q_0 = \pi/2$, $C=5$. Upper pane left: The I-V characteristics for various dimensionless temperatures $\tilde{T}$ marked in the labels. Lower pane:
Temperature-dependent plateux width for a set of fractional plateaux versus the temperature in units of the width obtained at current step $\Delta I = 0.01 \frac{2e \omega_0}{2 \pi}$. In contrast to Fig. \ref{fig:steps-T}, the curves are different for different fractions: there is no universal scaling.
  } \label{fig:steps-fT}
\end{figure}

Being encouraged with the success of the simple scaling hypothesis for the integer plateaux, we analyse the effect of finite temperature on the fractional plateaux in the intermediate regime $\tilde{C} \simeq 1$. The results are presented in Fig. \ref{fig:steps-fT}.  We observe there pronounced fractional plateaux at vanishing temperature, with the width up to $\simeq 0.1 V_\Sigma$. However, they vanish rather quickly, at the temperatures of two orders of magnitude lower than the integer ones. One could think that this is due to smaller width of the plateaux, so we check the simple scaling hypothesis plotting the temperature-dependent width in coordinates   ${\Delta V(T)}/{\Delta V(0)}$, $\tilde{T} V_\Sigma/\Delta V(0)$. We do not find the correspondence with the scaling of integer plateaux: the critical temperatures in units of the width are at least by a factor of 5 lower, and decrease with increasing the denominator. There is no scaling for different fractions, even for those with the same denominators. This may be explained by the fact that the shape of the effective energy barrier for fractional plateaux is different from that for integer plateaux, and is different from fraction to fraction. In conclusion, the fractional plateaux can only be observed at temperatures by two orders of magnitude lower than the integer ones.

\section{Conclusions} \label{sec:conclusion}

In conclusion, we propose to synchronize Bloch oscillations in a double phase-slip junction by modulating the gate voltage. This is advantageous in comparison with the bias voltage modulation since the a.c. signal does not produce extra dissipation that may kill the synchronization by overheating. We show that a.c. modulation gives rise to the pronounced  plateaux of quantized current of the width $V_\Sigma$ corresponding to the optimistic estimations for bias voltage modulation. 

We distinguish and investigate in detail three regimes corresponding to the ratio of the gate capacitance $C$ and effective junction capacitance $V_\Sigma/e$. The strong coupling regime $C \ll V_\Sigma/e$ is characterized by strong interference of the phase slips that is tuned by $q_0$, the charge induced by the d.c. part of the gate voltage. Well-developed plateaux are achieved at a.c. induced charge $\tilde{q}_g \simeq e$ corresponding to a.c. modulations 
The interference is suppressed in the opposite regime of weak coupling, $C \ll V_\Sigma/e$. The well-developed plateaux require bigger induced charge amplitudes $\tilde{q}_g \simeq C V_\Sigma$ but  smaller gate voltage amplitudes $\tilde{V}_g \simeq V_\Sigma$. Interestingly, well-developed fractional plateaux are developed in the intermediate regime  of  $C \simeq V_\Sigma/e$.

We investigate the effect of finite temperature on the smoothing of plateaux in all three regimes. The smoothing of integer plateaux is found to obey an empirical scaling law: the degree of smoothing is determined by the temperature in units of the plateau width. No such scaling was found for fractional plateaux that are only observable at temperatures by two orders of magnitude lower than the integer ones.    

To support open science and open software initiatives and to comply with institutional policies, we have published all relevant code and instructions for running it on the Zenodo repository\cite{BlochCode}. 

\acknowledgments

This research was supported by the European
Research Council (ERC) under the European Union's
Horizon 2020 research and innovation programme (grant
agreement No. 694272). We are grateful to S. E. de Graaf and O. V. Astafiev for several illuminating discussions that initiated this research.

\bibliography{bibliography}

\end{document}